\documentclass[amsmath, amssymb, 
twocolumn,
prb, 
]{revtex4-2}
\usepackage{braket}
\usepackage{graphicx}
\usepackage{color,xcolor}
\usepackage{ulem}
\usepackage{hyperref}
\usepackage{CJK}

\begin{document}
\begin{CJK*}{UTF8}{gbsn}

\title{Dynamical robustness of topological end states in nonreciprocal Su-Schrieffer-Heeger models with open boundary conditions}

\author{Li-Jun Lang (郎利君)} 
\email{ljlang@scnu.edu.cn}
\author{Yijiao Weng (翁益娇)}%
\author{Yunhui Zhang (张云辉)}%
\author{Enhong Cheng (成恩宏)}%
\author{Qixia Liang (梁绮霞)}%
\affiliation{Guangdong Provincial Key Laboratory of Quantum Engineering and Quantum Materials, School of Physics and Telecommunication Engineering, South China Normal University, Guangzhou 510006, China}
\affiliation{Guangdong-Hong Kong Joint Laboratory of Quantum Matter, Frontier Research Institute for Physics, South China Normal University, Guangzhou 510006, China}

\date{\today}

\begin{abstract}
For non-Hermitian quantum models, the dynamics is not apparently reflected by the static properties, e.g., the complex energy spectrum, because of the non-orthogonality of the right eigenvectors, the nonunitarity of the time evolution, the breakdown of the adiabatic theory, etc., but in experiments the time evolution of an initial state is commonly used. 
Here, we pay attention to the dynamics of an initial end state in nonreciprocal Su-Schrieffer-Heeger models under open boundary conditions, and we find that it is dynamically more robust than its Hermitian counterpart, because the non-Hermitian skin effect can suppress the part leaking to the bulk sites. 
To observe this, we propose a classical electric circuit with only a few passive inductors and capacitors, the mapping of which to the quantum model is established. 
This work explains how the non-Hermitian skin effect enhances the robustness of the topological end state, and it offers an easy way, via the classical electric circuit, of studying the nonreciprocal quantum dynamics, which may stimulate more dynamical studies of non-Hermitian models in other platforms.
\end{abstract}

\maketitle
\end{CJK*}

\section{Introduction}
In textbooks on quantum mechanics (e.g., Ref. \cite{110913-1}), the Hamiltonians must be Hermitian, required by the reality of the system's energy. Early in the 1990s, Hatano and Nelson discussed the effect of imaginary vector potentials on the delocalization transitions in non-Hermitian random quantum-mechanical problems \cite{181017-2,181017-3}; meanwhile, Bender and Boettcher \cite{101224-1} found that complex Hamiltonians with $\mathcal{PT}$ symmetry can also bear the real energy spectra when the $\mathcal{PT}$ symmetry of the eigenstates is not broken. Gradually, a theory of non-Hermitian quantum mechanics was established \cite{110707-1}, even for a general non-Hermitian Hamiltonian without the $\mathcal{PT}$ symmetry. Along with the development of the experimental techniques, non-Hermitian Hamiltonians can be engineered in many experimental labs, especially in topological photonics \cite{161004-1,Ozawa18}; also, the effective Hamiltonian for open quantum systems is in general non-Hermitian \cite{140717-1}. Therefore, the relevant theoretical problems naturally emerge as the concerns of theoretical physicists. Among them, exotic non-Hermitian topological phenomena \cite{Bergholtz2020} distinct from the Hermitian systems, e.g., $\mathcal{PT}$ symmetry breaking \cite{101224-1,Guo09,Peng14}, exceptional points \cite{Zhen15,Ding16,Doppler16,Xu16,Midya18,Jin2020,Zhang2020},  breakdown of the bulk-boundary correspondence of the Hermitian topological systems \cite{180720-9,Hu2011,Esaki2011,Zhu14,171012-1,180720-8,180720-7,Lieu18,180720-3,181022-1,180720-6,180720-4,Torres18,181016-1,181019-1,181221-2,181127-1,180924-1,190106-1,Wang2019a,Borgnia2020}, etc., have attracted considerable attention.
Recently, non-Hermitian effects have also been being considered in many-body problems \cite{200622-1,Kou2020,Shackleton2020,Matsumoto2019}.

Many works \cite{180720-9,Hu2011,Esaki2011,Zhu14,171012-1,180720-8,180720-7,Lieu18,180720-3,181022-1,180720-6,180720-4,Torres18,181016-1,181019-1,181221-2,181127-1,180924-1,190106-1,Wang2019a,Borgnia2020} have tried to define the topology in non-Hermitian systems by proposing various definitions of topological invariants, and to restore the bulk-boundary correspondence in non-Hermitian topological systems. Among them, based on the nonreciprocal Su-Schrieffer-Heeger (SSH) model, Yao {\it et al.} \cite{181022-1} found that the breakdown comes from the so-called ``non-Hermitian skin effect" under open boundary conditions (OBCs), and they established a ``non-Bloch" theory to restore the bulk-boundary correspondence for this chiral non-Hermitian model. This effect was theoretically studied recently in different contexts \cite{181221-1,Lee2019,Li2020,Yi2020,Lee2020,Hughes2020,Yang2020}, proposed to be realized on various platforms \cite{190802-1, Hofmann2019, Li2020a, Yoshida2020}, and observed in electric-circuit \cite{Thomal2020,Hofmann2020}, mechanical \cite{Ghatak2019}, photonic \cite{Xue2020,Weidemann2020}, and cold-atom \cite{Yan2020} experiments.

However, the bulk-boundary correspondence principle and the various definitions of the topological invariants have an implicit prerequisite, namely adiabatic theorem \cite{110608-1}, which breaks down in non-Hermitian dynamics due to the non-unitarity of the time evolution \cite{180924-2}. Therefore, many static topological properties, e.g., topological invariants and end/edge states, of non-Hermitian systems cannot be easily manipulated and observed. Compared with the static analysis, the dynamics is more straightforward for experiments \cite{Xue2020,Yan2020,Weidemann2020} to investigate the exotic properties of non-Hermitian systems, and it may also help to understand the dynamics of open quantum systems \cite{140717-1,Wang2019}. 

Moreover, although large energy gaps can protect the topological end/edge eigenstates of Hermitian models, part of the initial end/edge states will also leak into the bulk due to its superposition of the bulk eigenstates, affecting its dynamical robustness, which means how much of the initial end state can remain at the end in the time evolution.  If the nonreciprocity-induced non-Hermitian skin effect can suppress the bulk part, it would be useful to enhance the robustness of the topological end/edge states, which may be usefully exploited in amplifiers \cite{liang2013,Peano2016}, lasers \cite{171007-1, 180719-1,180719-2, 180419-3, 180402-1,180720-10, 180720-11,180409-1}, and other non-Hermitian photonic devices \cite{Leykam2020}.

In this paper, we focus on the dynamics of an initial end state in topologically different regimes of the nonreciprocal SSH model, and we find that the non-Hermitian skin effect indeed suppresses the bulk leakage and thus renders the time evolution of the initial end state more robust than its Hermitian counterpart.

To demonstrate this dynamical robustness, using a similar idea to that in Ref. \cite{190802-1}, we also propose an LC electric circuit to simulate the nonreciprocal SSH model. The electric circuit platform has been proved in recent years to be an easily-manipulated, low-cost, but powerful simulator for some topological phenomena \cite{170301-1,CHLee2018,Imhof2018,Hadad2018,171211-1,Yu2018,190307-1,Lu19,Garcia2019,Thomal2020,Hofmann2020}, but most of them focused on the driving schemes to study the resonance of the eigenstates \cite{Thomal2020,Hofmann2020}. Here, we use an initial end state to study its time evolution, which is much closer to the dynamical simulation of the quantum systems.
This work may stimulate more dynamical studies of non-Hermitian systems in other experimental platforms, such as photonics \cite{161004-1,Ozawa18,Leykam2020}, ultracold atoms \cite{Zhang2018}, and superconducting circuits \cite{Gu2017}.

\section{The nonreciprocal SSH model: recapitulation}
While the Hermitian SSH model \cite{130704-1} is a stereotypical one-dimensional (1D) topological model, the nonreciprocal variant, dubbed the nonreciprocal SSH model, is fundamental to understand the topology as well as the non-Hermitian skin effect in non-Hermitian systems.
In the second-quantization form, the Hamiltonian can be written as 
\begin{align}
 \hat{H}
   = \sum_n [\nu (\hat{a}^\dag_n \hat{b}_n + \hat{b}^\dag_n \hat{a}_n)+\kappa_2 \hat{a}^\dag_n \hat{b}_{n-1} + \kappa_1 \hat{b}^\dag_n \hat{a}_{n+1}], 
\label{Ham}
\end{align}
where $\hat{a}^{(\dag)}_n$ and $\hat{b}^{(\dag)}_n$ are the annihilation (creation) operators for the A- and B-sublattice sites, respectively, in the $n$th unit cell.
$\nu$ is the reciprocal intra-cell hopping amplitude, and $\kappa_{1,2}$ are the nonreciprocal inter-cell hopping amplitudes, leading to the non-Hermiticity of the system. All the parameters are real.

Although the conventional bulk-boundary correspondence principle for Hermitian topological systems breaks down in non-Hermitian ones \cite{181022-1,181016-1}, according to the ``non-Bloch" theory \cite{181022-1}, this nonreciprocal version of the SSH model also has non-Hermitian end eigenstates under OBCs with integer unit cells if parameters satisfy $|\kappa_1\kappa_2|>\nu^2$, which is continuable to the Hermitian SSH model with $\kappa_1=\kappa_2$. 
In addition, for $\kappa_1\ne \kappa_2$, all bulk eigenstates of $\hat{H}$ are localized near one boundary, depending on which module of $\kappa_{1,2}$ is larger. This phenomenon is called non-Hermitian skin effects, as shown in the insets at the top-left corners of Figs. \ref{fig2}(e) and \ref{fig2}(f).

\begin{figure}
  \includegraphics[width=0.9\linewidth]{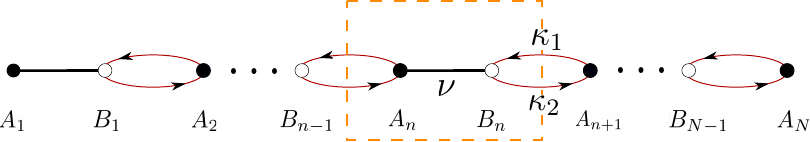}
  \caption{The sketch of the nonreciprocal SSH model under OBCs with $N$ unit cells, leaving the last one containing only the A-sublattice site to prevent the simultaneous appearence of the two end eigenstates. The solid (hollow) dots represent the A(B)-sublattice sites. The dashed (orange) square reflects a typical unit cell. The hopping amplitudes $\nu,\kappa_{1,2}$ are defined in the text. }
\label{fig1}
\end{figure}

To show the properties of the non-Hermitian end states in real space, for convenience we recast the Hamiltonian Eq. \eqref{Ham} under OBCs, as shown in Fig. \ref{fig1}, in the matrix form as $ \hat{H}  = \hat{\psi}^\dag H \hat{\psi}$, where $\hat{\psi}^\dag=(\hat{a}^\dag_1, \hat{b}^\dag_1,\cdots,\hat{a}^\dag_N)$ and $\hat{\psi}=(\hat{a}_1, \hat{b}_1,\cdots,\hat{a}_N)^T$ construct the vectors of basis $|\psi\rangle=\hat{\psi}^\dag|0\rangle$ and $\langle\psi|=\langle 0 | \hat{\psi}$, and
\begin{align}
H=\left(
\begin{array}{cccccc}
 0 & \nu \\
 \nu& 0 & \kappa_1   \\
  & \kappa_2   & 0 & \nu\\
  & &\ddots&\ddots &\ddots\\
  &&&\nu&0&\kappa_1 \\
  &&&&\kappa_2&0
\end{array}
\right)
\label{Ham_matrix}
\end{align}
forms a $(2N-1)\times (2N-1)$ Hamiltonian matrix of the coefficients with $N$ being the number of unit cells. 
Here, we keep only the A-sublattice site for the last unit cell to make sure that the two end eigenstates respectively localized at the two ends cannot appear simultaneously for all parameter cases, preventing the superposition of them for finite-size lattices.
Without loss of generality, $\nu$ is set positive as the energy unit, and we only consider the case of $\kappa_1\ge\kappa_2$, i.e., the non-Hermitian left-skin effect in the following unless otherwise stated.

\section{Quantum dynamics of end states}

\begin{figure}[htbp]
  \includegraphics[width=1\linewidth]{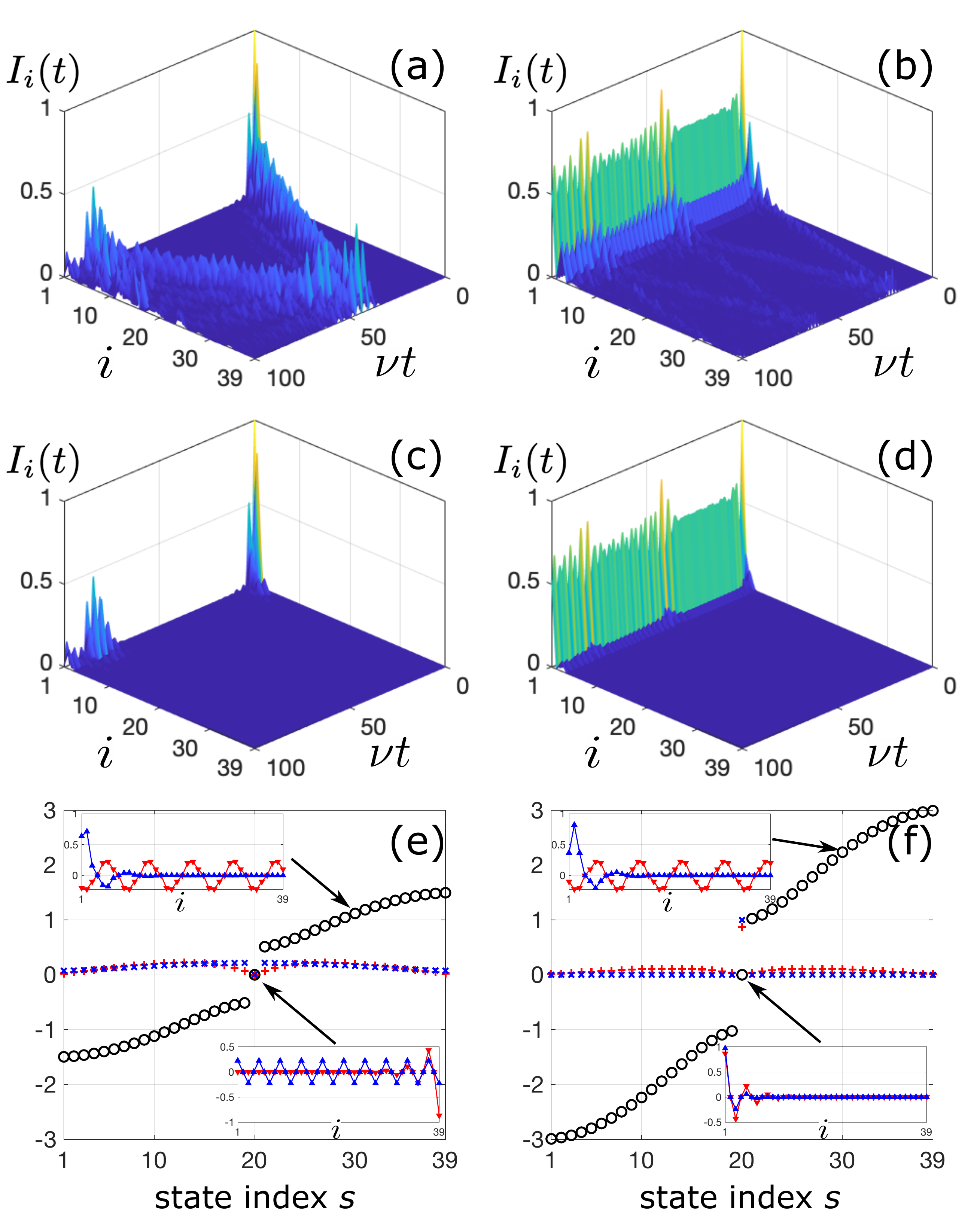}
  \caption{Time evolution of the relative intensity $I_i(t)=|\Psi_{i}(t)|^2/|\Psi_{1}(0)|^2$, given the initial state injected into the left-most site for Hermitian cases with $(\kappa_1,\kappa_2)/\nu=\textbf{(a) }(0.5,0.5)\text{ and {\bf(b)} }(2,2)$, and for nonreciprocal cases with $(\kappa_1,\kappa_2)/\nu=\textbf{(c) }(1,0.25)\text{ and {\bf(d)} }(4,1)$. 
{\bf(e)} The eigenenergy $E_s~(s=1,\cdots,2N-1)$ (black circles) and the weight $w_s$ (red pluses and blue crosses) of the initial end state, $\Psi(0)=(1,0,0,\cdots)^T$, on the $s$th right eigenvector $\psi^{(r)}_s$ of $H$, where $w_s=|\psi_s^{(l)}\Psi(0)|/(\sum_{s=1}^{2N-1}|\psi_s^{(l)}\Psi(0)|^2)^{1/2}$ with $\psi_s^{(l)}$ being the corresponding left eigenvector of $H$. 
The insets show the typical bulk and end right eigenvectors $\psi^{(r)}_s$ indicated by arrows.
The symbols of red pluses and down-pointing triangles are used for the Hermitian case with the parameters in (a), and blue crosses and up-pointing triangles for the nonreciprocal case in (c). Note that the Hamitonians of (a) and (c) have the same energy spectrum.
{\bf(f)} The same meaning as (e) but with the parameters used in (b) and (d) for Hermitian and nonreciprocal cases, respectively.
Here, we use $N=20$.} 
\label{fig2}
\end{figure}

It is well known that the end eigenstates of the Hermitian SSH model are topological, and thus are robust in dynamics. However, because an initial state in general contains a superposition of bulk eigenstates, according to the semiclassical theory for wave packets, the bulk part propagates into the bulk sites and is reflected by the two boundaries during evolution, and it will finally spread into the whole bulk after a sufficiently long time. 
So, the dynamical robustness of an initial end state for the Hermitian SSH model depends on the weight of the end eigenstate. 
Figures \ref{fig2}(a) and \ref{fig2}(b) show the typical dynamics of an initial end state for two Hermitian cases without and with the end eigenstate on the same end, respectively. The corresponding weights of the initial end state on all eigenstates are shown with red pluses in Figs. \ref{fig2}(e) and \ref{fig2}(f).

When we consider the nonreciprocal case, we may intuitively think that the (left-)skin effect should support the initial (left) end state staying at the (left) end in time, but Fig. \ref{fig2}(c), where there is no (left) end eigenstate [Fig. \ref{fig2}(e)], shows that the superposition of the (left-)skin bulk eigenstates of $\hat{H}$ still disperses the initial (left) end state into the bulk as in the Hermitian case, only with the magnitudes being suppressed.
When the (left) end right eigenstate exists, the initial (left) end state is dominated by it [Fig. \ref{fig2}(f)] and becomes dynamically robust, as shown in Fig. \ref{fig2}(d), where the superposition of the (left-)skin bulk eigenstates still leaks to the bulk as in Fig. \ref{fig2}(c), not influencing the dynamics of the end part. 

Compared with the Hermitian counterpart in Fig. \ref{fig2}(b), the initial end state is more robust in dynamics in the sense that the bulk parts are fully suppressed.
These counterintuitive phenomena demonstrate the novelty of the dynamics in non-Hermitian systems, which cannot be intuitively explained by the static energy spectrum and the superposition of the eigenstates as for the Hermitian ones.

To further understand these phenomena, we just need to notice the solution of the time-dependent Schr\"{o}dinger equation $i\partial_t \ket{\Psi(t)}=\hat{H}\ket{\Psi(t)}$ under OBCs (hereafter we set $\hbar=1$), i.e.,
\begin{equation}
\Psi(t)\equiv \langle \psi |\Psi(t)\rangle=e^{-iHt}\Psi(0)=Se^{-i\tilde{H}t}S^{-1}\Psi(0)
\label{evolution}
\end{equation}
under the basis $|\psi\rangle$ in real space, where 
\begin{align}
\tilde{H}=S^{-1}HS=\left(
\begin{array}{cccccc}
 0 & \nu \\
 \nu& 0 & \kappa   \\
  & \kappa & 0 & \nu\\
  & & \ddots &\ddots &\ddots\\  
  &&&\nu& 0&\kappa\\
 &&&&\kappa& 0
 \end{array}
\right)
\end{align}
is a $(2N-1)\times (2N-1)$ Hermitian matrix, mathematically similar to $H$ by an invertible matrix \cite{181022-1}
\begin{equation}
S=\mathrm{diag}(1,1,r^{-1},r^{-1},\cdots,r^{-(N-2)},r^{-(N-1)}),
\label{Smatrix}
\end{equation}
of which the exponentially decaying form is responsible for the non-Hermitian skin effect.
The parameters $\kappa=\sqrt{\kappa_1\kappa_2}>0$ and $r=\sqrt{\kappa_1/\kappa_2}\ge 1$, where we assume that $\kappa_{1,2}>0$. 

Focusing on Eq. \eqref{evolution}, with the aid of the similarity transformation, the final state $\Psi(t)$ of $H$ can be obtained from $\Psi(0)$ equivalently in three successive steps: (a) being scaled by $S^{-1}$; (b) evolving under the Hermitian matrix $\tilde{H}$; (c) being rescaled by $S$. 
Given an initial state confined in the left-most unit cell, $S^{-1}\Psi(0)$ in step (a) does not change it, preserving the signal only in the left-most unit cell. Then, the time evolution in step (b) just follows the Hermitian case as mentioned before: depending on the parameters, the signal propagates into the bulk for the topologically trivial case or keeping the dominant signal in the left-most unit cell with a small part propagating into the bulk for the topological case. However, different from the Hermitian case, the rescaling $S$ in step (c) eliminates exponentially the part leaking to the bulk.
This understanding can also naturally explain the periodic appearance of the peaks at the left end in Figs. \ref{fig2}(c) and \ref{fig2}(d), which come from the reflection of the bulk leaks by the right end of a finite-size lattice.
This phenomenon is also interesting, because it reflects the fact that although the bulk part is suppressed, it also evolves invisibly.

For $\kappa_1<\kappa_2$, which corresponds to the right-skin effect, the transformation matrix $S$ is exponentially amplifying, opposite to the left-skin case. Although the bulk leaks do not affect the end part, they will be amplifying to the right end and be finally dominant. This case is not what we care about, and in principle the Hamiltonian can always be engineered to make the skin-effect direction and the initial end state at the same end. Thus, in the following, we only focus on the case of $\kappa_1\ge\kappa_2$.

\section{Electric circuit's formalism of the nonreciprocal SSH model}

\begin{figure}
  \includegraphics[width=0.9\linewidth]{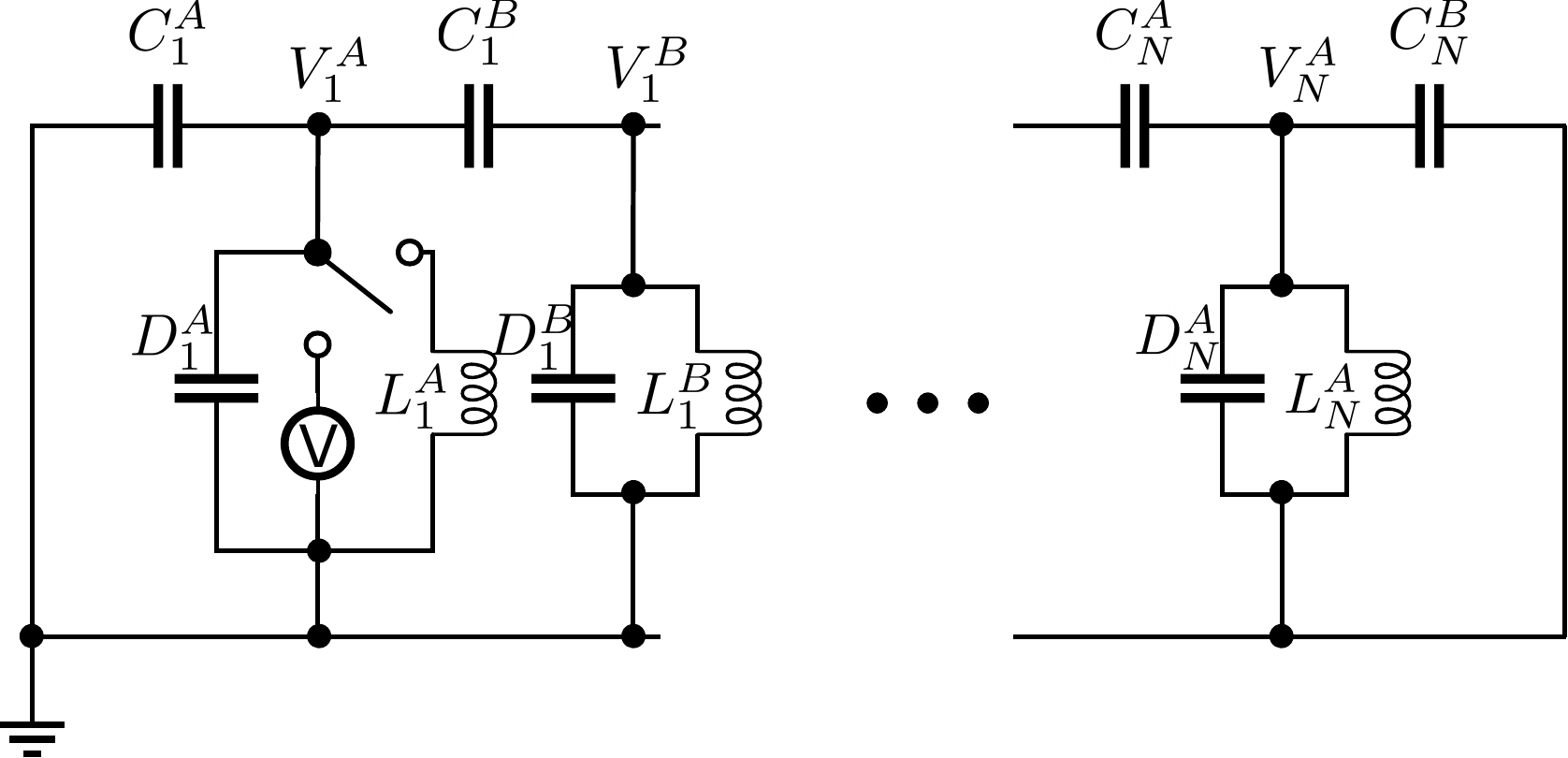}
  \caption{The scheme of the electric circuit to simulate the dynamics of the nonreciprocal SSH model, where the switch connects a DC voltage source to the left-most node for the initial state of the dynamics. To create the OBCs, both ends are grounded. Other parameters are defined in the text.}
\label{fig3}
\end{figure}

To simulate the fate of the initial end state in two topologically different phases of the nonreciprocal SSH model, we resort to LC electric circuits, as shown in Fig. \ref{fig3}, where we label the voltages of the A/B nodes in the $n$th unit cell as $V_n^{A/B}(t)$, the parameters $L_n^{A/B}$ as the corresponding inductances, and $C_n^{A/B}$ and $D_n^{A/B}$ as the capacitances.
 
We first do the static analysis, which corresponds to the switch in Fig. \ref{fig3} connecting to the inductor. Assuming $V_n^{A/B}(t)=V_n^{A/B}e^{i\omega t}$, according to the Kirchhoff's current law with the admittances of inductors and capacitors, i.e., $Y_L=1/i\omega L$ and $Y_C=i\omega C$, where $\omega$ is the angular frequency, we have the following set of equations:
\begin{align}
&\left(\Lambda_n^A-\frac{1}{\omega^2}\right)V_n^A =L_n^AC_n^A\,V_{n-1}^B+L_n^AC_n^B\,V_n^B, \notag\\
&\left(\Lambda_n^B-\frac{1}{\omega^2}\right)V_n^B =L_n^BC_n^B\,V_{n}^A+L_{n}^BC_{n+1}^A\,V_{n+1}^A.
\label{circuit_eq}
\end{align}
where $\Lambda_n^A=L_n^AD_n^A+L_n^AC_n^A+L_n^AC_n^B$ and $\Lambda_n^B=L_n^BD_n^B+L_n^BC_n^B+L_n^BC_{n+1}^A$.
We can also rewrite them in matrix form, $\mathcal{HV=EV}$, where $\mathcal{V}=(V_1^A,V_1^B,\cdots,V_N^A)^T$,
\begin{equation}
\mathcal{H}=\left(
\begin{array}{ccccc}
 0 & L_1^AC_1^B \\
 L_1^BC_1^B& 0 & L_1^BC_{2}^A   \\
  &\ddots&\ddots &\ddots\\
    &&L_N^AC_N^A&0
\end{array}
\right), 
\label{circuit_matrix}
\end{equation}
and
\begin{equation}
\mathcal{E}=\Lambda-\frac{1}{\omega^2}I=\mathrm{diag}\left(\Lambda_1^A,\Lambda_1^B,\cdots,\Lambda_N^A
\right)-\frac{1}{\omega^2}I,
\label{eigenvalue}
\end{equation}
where $I$ is the identity matrix.

To map this set of equations to the nonreciprocal SSH model, we just regard $\mathcal{H}$ as the Hamiltonian matrix $H$, and $\mathcal{V}$ and $\mathcal{E}$ respectively as its right eigenvector and eigenenergy. Thus, $\mathcal{HV=EV}$ can simulate the time-independent Schr\"{o}dinger equation of the nonreciprocal SSH model if $\mathcal{E}$ and thus $\Lambda$ are proportional to the identity matrix.

To establish the mapping between the parameters of two Hamiltonian systems, Eqs. \eqref{Ham_matrix} and \eqref{circuit_matrix}, we must have the following relations (in dimensionless way):
\begin{align}
&\frac{L_n^AC_n^B}{LC}=\frac{L_n^BC_n^B}{LC}=1,  \notag\\
&\frac{L_n^BC_{n+1}^A}{LC}=\frac{\kappa_1}{\nu}, ~~~~
\frac{L_n^AC_n^A}{LC}=\frac{\kappa_2}{\nu},  \notag\\
&\frac{\Lambda_n^A}{LC}=\frac{\Lambda_n^B}{LC}=\lambda,~~~~(n=1,\cdots,N),
\label{condition}
\end{align}
where $\lambda$ is a dimensionless constant taken as needed, and $L$ and $C$ are just reference parameters of inductors and capacitors, respectively, taken for practical need.
The solution for the inductors and capacitors in the LC circuit is as follows:
\begin{align}
&C_n^A=C\left(\frac{\kappa_1}{\kappa_2}\right)^{n-1}, ~~~~~~
C_n^B=C_n^A\frac{\nu}{\kappa_2}, \notag\\
&D_n^A=C_n^A\frac{\lambda\nu-\nu-\kappa_2}{\kappa_2}, ~~~
D_n^B=C_n^A\frac{\lambda\nu-\nu-\kappa_1}{\kappa_2}, \notag\\
&L_n^A=L_n^B=L\left(\frac{\kappa_2}{\kappa_1}\right)^{n-1}\frac{\kappa_2}{\nu},~~~~(n=1,\cdots,N).
\label{relation}
\end{align}
Note that the inductances and the capacitances must be non-negative, which can be realized by setting $L,C>0$ and $\lambda\ge 1+\max(\kappa_1/\nu,\kappa_2/\nu)$, accompanied by the model parameters $\kappa_{1,2},\nu>0$ mentioned before.  
For convenience, we can set $\lambda=1+\kappa_1/\nu$ for the case of $\kappa_1\ge\kappa_2$ to remove the inductances $D_n^B$ in the circuit, i.e., $D_n^B=0$, and thus, $D_n^A=C_n^A(\kappa_1-\kappa_2)/\kappa_2.$
In these settings, we have the effective Hamiltonian matrix $\mathcal{H}=(LC/\nu)H$ and the effective eigenenergy $\mathcal{E}=(\lambda LC-1/\omega^2)I.$

We can also use another circuit scheme by just exchanging inductors and capacitors. The results can be easily obtained by the duality replacement: $\omega^2\rightarrow 1/\omega^2, ~C_n^{A/B}\leftrightarrow 1/L_n^{A/B},~D_n^{A/B}\rightarrow 1/l_n^{A/B}$ in Eqs. \eqref{circuit_eq}, \eqref{circuit_matrix}, and \eqref{relation}, where $l_n^{A/B}$ is another label of inductors.

\section{Electric circuit's dynamical simulation}
For the dynamical simulation, we need to consider the original differential equation, i.e.,
\begin{equation}
(\mathcal{H}-\Lambda)\ddot{\mathcal{V}}(t)-\mathcal{V}(t)=0,
\label{t-equation}
\end{equation}
where $\mathcal{H}$ and $\Lambda$ are defined generally in Eqs. \eqref{circuit_matrix} and \eqref{eigenvalue}, and $\mathcal{V}(t)=[V_1^A(t),V_1^B(t),\cdots,V_N^A(t)]^T$. The dot in $\dot{\mathcal{V}}$ means the derivative with respect to time, i.e., $d\mathcal{V}/dt$.
Under the substitutions, Eqs. \eqref{condition} or \eqref{relation}, the general solution reads
\begin{equation}
\mathcal{V}(t)=\sum_{s=1}^{2N-1}(\alpha_s\cos\omega_s t+\beta_s\sin\omega_s t)\cdot\mathcal{V}_s,
\label{gel_sol}
\end{equation}
where $\mathcal{V}_s$ is the $s$th right eigenvector of $\mathcal{H}$, and $\omega_s$ is the corresponding eigen angular frequency of the circuit. $\{\alpha_s,\beta_s\}$ are a set of complex coefficients to be determined by the initial conditions, coming from the second order derivatives with respect to time in Eq. \eqref{t-equation}. 
The detailed derivation can be referred to in the Appendix. 
The form of the superposition coefficients in Eq. \eqref{gel_sol} is different from our familiar form of the solution of the time-dependent Schr\"{o}dinger equation, but it is qualitatively the same due to the same fact that the solution is the superposition of the Hamiltonian's eigenvectors.

To simulate the previously mentioned dynamics of the two topologically different phases, we adopt the LC circuit in Fig. \ref{fig3}. 
For the following selected values of parameters $(\kappa_1/\nu,\kappa_2/\nu,\lambda)$, we set the reference parameters in Eq. \eqref{relation} as $L=1$mH and $C=100$pF, and the number of unit cells $N=5$, which ensures that all values of circuit elements drop into the available regime in the realistic experiment. The typical oscillating angular frequency of the circuit is $\omega_0=1/\sqrt{LC}$ in the order of MHz.

For the topological phase, we choose $(\kappa_1/\nu,\kappa_2/\nu,\lambda)=(4,1,5)$ to simulate the dynamics of the nonreciprocal SSH model, corresponding to $C_n^A=C_n^B=100\mathrm{pF}\sim25.6 \mathrm{nF}$, $D_n^A=300\mathrm{pF}\sim76.8 \mathrm{nF}$, $D_n^B=0$, and $L_n^A=L_n^B=1\mathrm{mH}\sim 3.9\mathrm{\mu H}$; as comparison, we also choose $(\kappa_1/\nu,\kappa_2/\nu,\lambda)=(2,2,5)$ for the Hermitian counterpart, corresponding to $D_n^A=D_n^B=C_n^A=100\mathrm{pF}$, $C_n^B=50\mathrm{pF}$, and $L_n^A=L_n^B=2\mathrm{mH}$. Here, we use $n=1,\cdots,5$. These two cases are said to be topological and have the topological end right eigenstate on the left end because $|\kappa_1\kappa_2/\nu^2|=4>1$.

For the topologically trivial phase, we choose $(\kappa_1/\nu,\kappa_2/\nu,\lambda)=(1,0.25,2)$, corresponding to $C_n^A=100\mathrm{pF}\sim25.6 \mathrm{nF}$, $C_n^B=400\mathrm{pF}\sim102.4 \mathrm{nF}$, $D_n^A=300\mathrm{pF}\sim76.8 \mathrm{nF}$, $D_n^B=0$, and $L_n^A=L_n^B=250\sim 1\mathrm{\mu H}$; as a comparison, we also choose $(\kappa_1/\nu,\kappa_2/\nu,\lambda)=(0.5,0.5,2)$ for the Hermitian counterpart, corresponding to $D_n^A=D_n^B=C_n^A=100\mathrm{pF}$, $C_n^B=200\mathrm{pF}$, and $L_n^A=L_n^B=500\mathrm{\mu H}$. They are topologically trivial and have no topological end right eigenstate on the left end because $|\kappa_1\kappa_2/\nu^2|=0.25<1$.

The initial left end state can be prepared by connecting the switch to the DC voltage source, and after a sufficient time, only capacitors $C_1^A$ and $C_1^B$ are charged, which renders only $V_1^A(0)=V_0$ with $V_0$ being the value of the DC voltage source, and thus determines the values of $\{\alpha_s\}$.
By changing the switch to the inductor in the first unit cell at time $t=0$, the zero transient current renders $\dot{V}_n^{A/B}=0$ for all $n$, and thus vanishes $\{\beta_s\}$. The determination of $\{\alpha_s,\beta_s\}$ can be referred to in the Appendix. Then, $\mathcal{V}(t)$ of any time can be simulated according to Eq. \eqref{gel_sol}. 

\begin{figure}
  \includegraphics[width=1\linewidth]{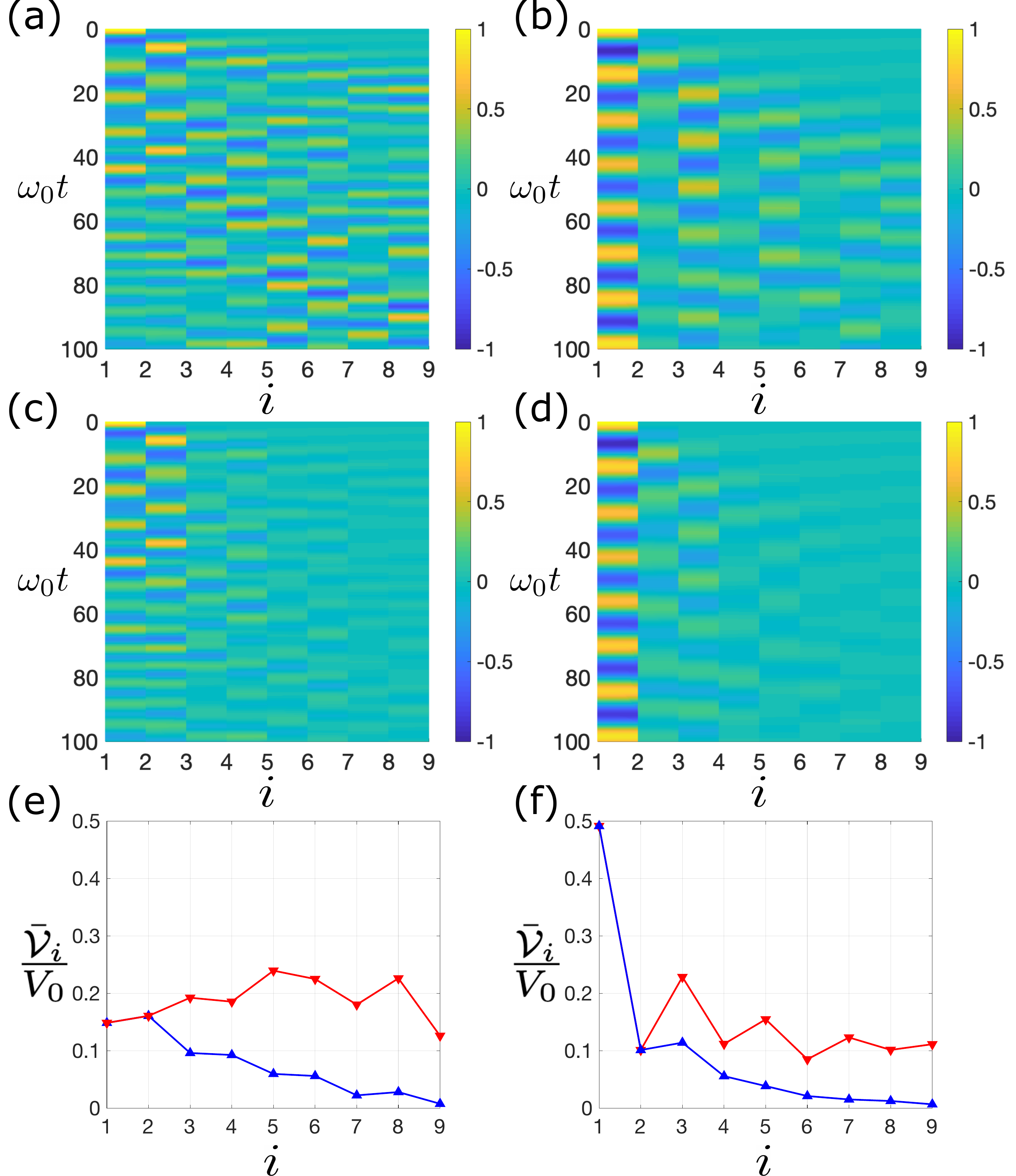}
  \caption{Time evolution of the relative voltages, $\mathcal{V}_i(t)/V_0$, in the circuit of Fig. \ref{fig3}, given the initial condition by the switch changing described in the text, for the Hermitian SSH circuits with {\bf(a)} $(\kappa_1/\nu,\kappa_2/\nu,\lambda)=(0.5,0.5,2)$ and {\bf(b)} $(2,2,5)$, and for the nonreciprocal SSH circuits with {\bf(c)} $(\kappa_1/\nu,\kappa_2/\nu,\lambda)=(1,0.25,2)$ and {\bf(d)} $(4,1,5)$. 
{\bf(e,f)} The average voltages, $\bar{\mathcal{V}}_i(t)/V_0$ (defined in the text), at $\omega_0 t=100$ for the cases (a,c) without and (b,d) with the left-end right eigenstates, respectively, where the Hermitian ones are labeled by the red lines with down-pointing triangles and the nonreciprocal ones by the blue lines with up-pointing triangles.}  
\label{fig4}
\end{figure}

Figure \ref{fig4} shows the simulation results. 
As expected, compared with their Hermitian counterparts [Fig. \ref{fig4}(a) and \ref{fig4}(b)], we can clearly see that the nonreciprocity of both topologically trivial and nontrivial cases can suppress the bulk leakage [Fig. \ref{fig4}(c) and \ref{fig4}(d)]. Thus, the initial end state for the topological case becomes more robust in dynamics [Fig. \ref{fig4}(d)]. 
Figure \ref{fig4}(c) shows dynamically the non-Hermitian skin effect in the topologically trivial circuit, where the nonreciprocal state has the bulk part suppressed, in contrast to the Hermitian counterpart, which spreads uniformly in the whole circuit. 
Figure \ref{fig4}(f) just confirms that the end state in the topological case for the nonreciprocal SSH circuit is dynamically more localized than the Hermitian counterpart. 

To quantitatively reflect the extent of localization, we calculate the average inverse participation ratio (aIPR), defined as
\begin{eqnarray}
\mathrm{aIPR}&=&\frac{\sum_{i=1}^{2N-1}[\bar{\mathcal{V}}_i(t)]^4}{\sum_{i=1}^{2N-1}[\bar{\mathcal{V}}_i(t)]^2},
\end{eqnarray}
where
\begin{eqnarray}
\bar{\mathcal{V}}_i(t)&=&\frac{2}{t}\int_{t/2}^{t}|\mathcal{V}_i(\tau)|d\tau.
\end{eqnarray}
$\mathrm{aIPR}=1$ means that the averaged voltage distribution in the time interval $(t/2,t)$ is totally localized at one circuit node, while $\mathrm{aIPR}=1/(2N-1)$ means it is uniformly extended in the whole circuit. 
As the quantitative verification of the previous judgment, the aIPRs are equal to 0.1262 (red down-pointing triangle line) and 0.2452 (blue up-pointing triangle line) for the curves in Fig. \ref{fig4}(e), and 0.4194 (red down-pointing triangle line) and 0.8017 (blue up-pointing triangle line) in Fig. \ref{fig4}(f).

Actually, the dynamical behaviors in Fig. \ref{fig4} can also be understood using the circuit's version of Eq. \eqref{evolution}.
To achieve this, we rewrite Eq. \eqref{t-equation} into the form of a first order differential equation as follows:
\begin{align}
\left(
\begin{array}{c}
 \dot{\mathcal{V}}(t) \\
 \dot{\mathcal{W}}(t)   \\
\end{array}
\right)=
\left(
\begin{array}{cc}
 0&I \\
 (\mathcal{H}-\Lambda)^{-1} & 0  \\
\end{array}
\right)
\left(
\begin{array}{c}
 \mathcal{V}(t) \\
 \mathcal{W}(t)   \\
\end{array}
\right)
\end{align}
by introducing $\mathcal{W}(t)=\dot{\mathcal{V}}(t)$ as new variables. And now, it is more like the time-dependent Schr\"{o}dinger equation, and the solution can be written as
\begin{align}
\left(
\begin{array}{c}
 \mathcal{V}(t) \\
 \mathcal{W}(t)   \\
\end{array}
\right)
=&
\exp\left[
\left(
\begin{array}{cc}
 0&I \\
 (\mathcal{H}-\Lambda)^{-1} & 0  \\
\end{array}
\right)t\right]
\left(
\begin{array}{c}
 \mathcal{V}(0) \\
 \mathcal{W}(0)   \\
\end{array}
\right) \notag\\
=&
\left(
\begin{array}{cc}
 S & 0\\
  0 & S  \\
\end{array}
\right)
\exp\left[
\left(
\begin{array}{cc}
 0&I \\
 (\tilde{\mathcal{H}}-\Lambda)^{-1} & 0  \\
\end{array}
\right)t\right]\notag \\
&\times\left(
\begin{array}{cc}
 S^{-1} & 0 \\
 0 & S^{-1}  \\
\end{array}
\right) 
\left(
\begin{array}{c}
 \mathcal{V}(0) \\
 \mathcal{W}(0)   \\
\end{array}
\right),
\end{align}
where $\tilde{\mathcal{H}}=S^{-1}\mathcal{H}S$ is the Hermitian counterpart, with $S$ being defined in Eq. \eqref{Smatrix}. This is qualitatively the same as Eq. \eqref{evolution}, and thus, the same explanation also applies to the circuit SSH systems.

\section{Conclusion and discussion}
The nonreciprocal SSH model is fundamental to the understanding of the non-Hermitian physics. In this paper, we transfer the focus from the static properties to the direct dynamics, showing that the non-Hermitian skin effect of this model under OBCs can enhance the robustness of the dynamics of the topological end states with a succinct explanation. Then, we propose an electric circuit with only a few passive elements of linear inductors and capacitors to simulate the nonreciprocal SSH model, clearly demonstrating the advantage in dynamics over the Hermitian counterparts for the topological end states. 

In this paper, we restrict the model parameters as $\kappa_{1.2},\nu>0$, which can be easily extended to $(\kappa_{1}/\nu)(\kappa_{2}/\nu)>0$, showing the same dynamical properties, because the similarity transformation $S$ still works to map it back to a Hermitian counterpart. However, for $(\kappa_{1}/\nu)(\kappa_{2}/\nu)<0$, the eigenenergy or eigen angular frequency can in general be complex, and thus, the amplitude will be amplifying or decaying in time; furthermore, the circuit scheme with Eq. \eqref{relation} cannot simulate this case due to the positivity of the circuit elements, which is beyond the scope of our discussion.

For experiments, the inductors and capacitors do not have to be selected with the exact values in the main text, because the non-Hermitian skin effect here just depends on the increasing or decreasing of the values, not the exact power law in Eq. \eqref{relation}. Therefore, the small intrinsic inductances and capacitances of the circuit cannot qualitatively affect the results either, nor do the small resistances of the connecting wires in the circuit, which only quantitatively decays the amplitudes in a long time.

For the initialization of the time evolution in electric-circuit experiments,  in principle the switching time must be much smaller than the discharging time of the capacitors to ensure the little change of the initial state at $t=0$.
Alternatively, we may consider the driving scheme to investigate the robustness of the nonreciprocal SSH end state as in Ref. \cite{190802-1}, but that is beyond the scope of this paper.

\begin{acknowledgments}
L.-J. L. was supported by National Natural Science Foundation of China (No. 11904109), Guangdong Basic and Applied Basic Research Foundation (No. 2019A1515111101), Science and Technology Program of Guangzhou (No. 2019050001), and the startup fund from South China Normal University.
\end{acknowledgments}
\appendix
\section*{Dynamical solution of the electric circuit}\label{a-sol}
Under the substitutions, Eqs. \eqref{condition} or \eqref{relation}, where note that $\Lambda=(\lambda LC) I$ is proportional to the identity matrix, we can make the following eigenvalue decomposition from Eqs. \eqref{circuit_matrix} and \eqref{eigenvalue}:
\begin{eqnarray}
M^{-1}\mathcal{H}M=\mathcal{E}=\Lambda-(\Omega^2)^{-1},
\label{similar}
\end{eqnarray}
where the columns of $M$ are just the $2N-1$ right eigenvectors of $\mathcal{H}$, i.e.,  $\{\mathcal{V}_s\}$, and $\Omega^2=\mathrm{diag}(\omega_1^2,\cdots,\omega_{2N-1}^2)$ is the set of corresponding eigen angular frequencies of the circuit, $\{\omega_s\}$ .

For the differential equation \eqref{t-equation}, applying the similarity transformation, Eq. \eqref{similar}, to both sides, we have
\begin{eqnarray}
&&M^{-1}(\mathcal{H}-\Lambda)M[M^{-1}\ddot{\mathcal{V}}(t)]-[M^{-1}\mathcal{V}(t)]=0 \notag\\
&\Rightarrow&-(\Omega^2)^{-1}[M^{-1}\ddot{\mathcal{V}}(t)]-M^{-1}\mathcal{V}(t)=0 \notag\\
&\Rightarrow&[M^{-1}\ddot{\mathcal{V}}(t)]+(\Omega^2)[M^{-1}\mathcal{V}(t)]=0.
\end{eqnarray}
Because $\Omega^2$ is a diagonal matrix, we can consider the solution for each element independently, i.e., 
\begin{eqnarray}
[M^{-1}\mathcal{V}(t)]_s&=&C_se^{i\omega_s t}+D_se^{-i\omega_nt}\notag\\
&=&\alpha_s\cos\omega_s t+\beta_s\sin\omega_s t,
\end{eqnarray}
where $\{C_s,D_s\}$ and $\{\alpha_s=C_s+D_s,\beta_s=i(C_s-D_s)\}$ are two sets of complex coefficients to be determined by the initial conditions.
Then, the finial solution reads
\begin{eqnarray}
\mathcal{V}(t)=MT(t),
\end{eqnarray}
where $T(t)$ is an $(2N-1)\times 1$ coefficient vector with entries $\{C_se^{i\omega_s t}+D_se^{-i\omega_st}\}$ or $\{\alpha_s\cos\omega_s t+\beta_s\sin\omega_s t\}$.
Then we get the formula Eq. \eqref{gel_sol}.

Given the initial conditions, 
\begin{eqnarray}
(\alpha_1,\cdots,\alpha_{2N-1})^T&=&T(0)=M^{-1}\mathcal{V}(0),\notag \\
(\beta_1\omega_1,\cdots,\beta_{2N-1}\omega_{2N-1})^T&=&\dot{T}(0)=M^{-1}\dot{\mathcal{V}}(0),
\end{eqnarray}
the coefficients, say $\{\alpha_s,\beta_s\}$, can be determined, and thus, $\mathcal{V}(t)$ at any time is obtained.

\bibliography{ref}
\bibliographystyle{apsrev4-2}

\end{document}